%% file: main.tex
\documentclass[sigconf,authorversion,nonacm]{acmart}

\newif\ifanonymousversion
\anonymousversionfalse

\usepackage[utf8]{inputenc}
\usepackage{xcolor}
\usepackage{soul}
\usepackage{balance}
\usepackage{url}
\usepackage{graphicx}
\usepackage{caption}
\usepackage{subcaption}
\usepackage{quantikz}

\newcommand{\nsf}[1]{\href{https://www.nsf.gov/awardsearch/showAward?AWD_ID=#1}{#1}}

\begin{document}

\title[Short Paper: Device- and Locality-Specific Fingerprinting of Shared NISQ Quantum Computers]{Short Paper: Device- and Locality-Specific Fingerprinting\\ of Shared NISQ Quantum Computers\vspace{1em}}

\ifanonymousversion

\author{Anonymous Author(s)}

\else

\author{Allen Mi}
\affiliation{%
  \institution{Yale University}
}
\email{allen.mi@yale.edu}

\author{Shuwen Deng}
\affiliation{%
  \institution{Yale University}
}
\email{shuwen.deng@yale.edu}

\author{Jakub Szefer}
\affiliation{%
  \institution{Yale University}
}
\email{jakub.szefer@yale.edu}

\renewcommand{\shortauthors}{Allen Mi, Shuwen Deng, and Jakub Szefer}

\fi

\input{abstract}

\maketitle


\input{introduction}

\input{background}

\input{methodology}

\input{evaluation_setup}

\input{results}

\input{conclusion}

\input{acknowledgement}

\balance

\bibliographystyle{ACM-Reference-Format}
\bibliography{main}

\end{document}

%% file: abstract.tex
\begin{abstract}

Fingerprinting of quantum computer devices is a new threat that poses a challenge to shared, cloud-based quantum computers. Fingerprinting can allow adversaries to map quantum computer infrastructures, uniquely identify cloud-based devices which otherwise have no public identifiers, and it can assist other adversarial attacks. This work shows idle tomography-based fingerprinting method based on crosstalk-induced errors in NISQ quantum computers. The device- and locality-specific fingerprinting results show prediction accuracy values of $99.1\%$ and $95.3\%$, respectively. 

\end{abstract}

%% file: introduction.tex
\section{Introduction}

Today's quantum computers are commonly called Noisy Intermediate-Scale Quantum (NISQ) quantum computers~\cite{preskill2018quantum}. NISQ quantum computers are small, but have promising applications in optimization, chemistry, and other important areas~\cite{lanyon2010towards,jones1998implementation,mermin2007quantum}. Further, quantum computing hardware keeps evolving at a fast pace, and $1000$-qubit quantum computers are projected to come online in near future~\cite{gambetta2020ibm}. As this increasing number of qubits are available, ideas for multi-programming and shared quantum computers have emerged~\cite{das2019case}. Instead of allocating all qubits to a single task or user, researchers have been exploring how the computers can be shared between  different users or tasks. Sharing of the quantum computers can improve the utilization of the resources and eventually lower costs for users. But it comes at a security cost.

\subsection{Cloud-based Quantum Computers}

There is now a growing interest in, and practical deployments of, cloud-based quantum computers, also called Quantum as a Service (QaaS). Among others, IBM is providing free access to its quantum processors and simulators through IBM Q service. Other cloud-based vendors providing quantum computer access today include Amazon Braket and Microsoft Azure.

Cloud-based quantum computing opens up many new opportunities, not just for renting single-user quantum computers, as is done today, e.g., through IBM Q, but for multi-programmed and shared quantum computers~\cite{das2019case}. With cloud-based access, the provider can decide which quantum computer to schedule the programs on, or it can put two or more programs (users) on the same computer if the resources allow. Time sharing of resources is not possible in quantum computers yet, but spatial sharing of qubits is possible.

\subsection{Security Challenges of Cloud-based Quantum Computer Architectures}

Cloud-based quantum computers are vulnerable to many threats not present in in-house
uses of quantum computers. For example, the remote users can be malicious and try to learn the infrastructure, harm the infrastructure, attack other users, or leak information from other users.

On the attack side, information leakage~\cite{phalak2021quantum} or attempts at interference with other users~\cite{ash2020analysis}, have now been proposed and demonstrated in emulated multi-tenant setting. Others have shown also that when malicious users share the same quantum computer as the victim, they can try to exploit crosstalk to perform fault injection in a quantum machine learning classifier, e.g., to increase the probability of misclassification~\cite{ash2020analysis}.

These attacks implicitly assume that the attacker is able to locate himself or herself on a specific quantum computer (i.e., device), or within the quantum computer (i.e., locality). Early work~\cite{phalak2021quantum} has proposed a very simple Quantum Physically Unclonable Functions (QuPUFs) design based on readout error or one-qubit gate error and considered two older IBM Q machines.

This work advances the state of the art with a new device fingerprinting approach for both device-specific and locality-specific fingerprinting based on idle tomography, showing superior accuracy and evaluation on $9$ current machines and dozens of possible subgraph embeddings within these devices.

\subsection{Contributions}

The contributions of this work are:

\begin{itemize}
\item Development of a crosstalk-based fingerprinting approach for NISQ quantum computers.
\item Demonstration of reliable device-specific and location-specific fingerprinting for quantum computers.
\item Evaluation on 9 IBM Q superconducting machines with various qubit sizes and topologies.
\end{itemize}

%% file: background.tex
\section{Background}

This section introduces background on crosstalk and idle tomography, in-depth details of quantum computers and their architectures are available from existing work, e.g.,~\cite{mermin2007quantum}.

\subsection{Noise and Crosstalk}

Noise in quantum computers can be attributed to gate errors, decoherence errors, readout errors, and crosstalk errors. Gate errors can affect single-qubit gates such as the Hadamard gate and two-qubit gates such as the \textsc{CNOT} gate. Decoherence errors are due to the interaction of the qubits with the surrounding environment, as a result of which their state is lost or modified. Readout errors are errors that occur in measurement operations that affect the readout probabilities. Crosstalk errors result as gate operations on one or two qubits (depending on the gate type) affect other, nearby qubits or gates. The crosstalk can be qubit-to-qubit, coupling-to-qubit, qubit-to-coupling, or coupling-to-coupling. As this work shows, crosstalk is a feature of NISQ quantum computer hardware that allows adversarial threats.

\subsection{Measuring Crosstalk and Idle Tomography}

Simultaneous Randomized Benchmarking (SRB)~\cite{gambetta2012characterization} or Idle Tomography (IDT)~\cite{blume2019idle} can be used to measure crosstalk. They both aim to quantify crosstalk in terms of error rates.  IDT has been recently proposed by Sandia Labs~\cite{blume2019idle}, it uses a comparably small number of circuits and relatively short circuits, and is the method selected in this work due to its simplicity and~effectiveness.

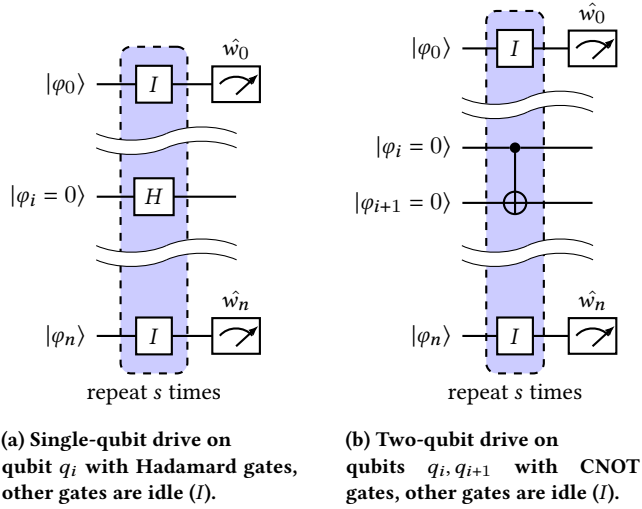
\begin{figure}
    \centering
    \begin{subfigure}[b]{0.46\linewidth}
        \begin{quantikz}
            \lstick{\ket{\varphi_0}} & \gate{I}\gategroup[wires=5,steps=1,style={dashed, rounded corners,fill=blue!20, inner xsep=2pt}, background,label style={label position=below,anchor= north,yshift=-0.2cm}]{{repeat $s$ times}} & \meter{$\hat{w_0}$} \\
            \wave&& \\
            \lstick{\ket{\varphi_i = 0}} & \gate{H} & \qw  \\
            \wave&& \\
            \lstick{\ket{\varphi_n}} & \gate{I} & \meter{$\hat{w_n}$}
        \end{quantikz}
        \caption{\small Single-qubit drive on \\ qubit $q_i$ with Hadamard gates, other gates are idle ($I$).}
    \end{subfigure}
    \hfill
    \begin{subfigure}[b]{0.46\linewidth}
        \begin{quantikz}
            \lstick{\ket{\varphi_0}} & \gate{I}\gategroup[wires=6,steps=1,style={dashed, rounded corners,fill=blue!20, inner xsep=2pt}, background,label style={label position=below,anchor= north,yshift=-0.2cm}]{{repeat $s$ times}} & \meter{$\hat{w_0}$} \\
            \wave&& \\
            \lstick{\ket{\varphi_i = 0}} & \ctrl{1} & \qw  \\
            \lstick{\ket{\varphi_{i+1} = 0}} & \targ{} & \qw \\
            \wave&& \\
            \lstick{\ket{\varphi_n}} & \gate{I} & \meter{$\hat{w_n}$}
        \end{quantikz}
        \caption{\small Two-qubit drive on \\ qubits $q_i, q_{i+1}$ with \textsc{CNOT} gates, other gates are idle ($I$).}
    \end{subfigure}
    \caption{\small Circuit schematic of idle tomography circuits with single- and two-qubit drive. Each spectator qubit $q_j$ is initialized in a Pauli basis state $\ket{\varphi_j}$ and measured with respect to a Pauli basis $\hat{w_j} \in \{\hat{x}, \hat{y}, \hat{z}\}$. The idle length $s$ determines the number of times the drive gates are repeated.}
    \label{fig:idt}
\end{figure}

The principle for IDT is to characterize the error accumulated by idle qubits over time. IDT is effective at measuring how the influence of gate operations propagates to other qubits via crosstalk. Figure \ref{fig:idt} displays a typical IDT setup, in which one or two qubits are selected as \textit{drive qubits}. These qubits are prepared in the $\ket{0}$ state in the logical basis. The rest of the qubits are \textit{spectator qubits}. Each spectator qubit is prepared in one of the Pauli bases $\hat{x}$, $\hat{y}$, or $\hat{z}$. After preparation, gate operations commence on the drive qubits. Commonly, the Hadamard gate $H$ and the controlled-not gate $\textsc{CNOT}$ are used for single- and two-qubit drive cases respectively. In the meantime, spectator qubits are kept at idle. We define the \textit{idle length} as the number of times the gate operations on the drive qubits are repeated. Finally, we measure each spectator qubit in one of the Pauli bases and output the measurement results, which are further used for characterizing the Hamiltonian, stochastic and affine error~channels. In addition, control-group experiments can also be utilized to characterize error channels due to ambient effects. These experiments take all qubits as spectators, and use the same idle length values as in the drive cases.

Given a combination of a driver gate and driver qubit(s), a complete set of idle tomography circuits on $n'$ spectator qubits is enumerated from a combination of the following parameters:
\begin{itemize}
    \item Initialization of each spectator qubit in a Pauli basis.
    \item Measurement of each spectator qubit in a Pauli basis.
    \item Idle lengths $s \in S$, a set of idle length values.
\end{itemize}
In practice, we take a subset of this complete set by limiting $S$ and the values each parameter can take. These decisions depend on the size and topology of the target device, as well as the desired granularity of error characterization.

%% file: methodology.tex
\section{Methodology}

This section is devoted to demonstrate the fingerprinting approach. We outline the threat model in Subsection \ref{subsec:threat_model} and discuss each component of the fingerprinting process in the subsequent subsections.

\begin{figure*}[t!]
  \includegraphics[width=0.95\linewidth]{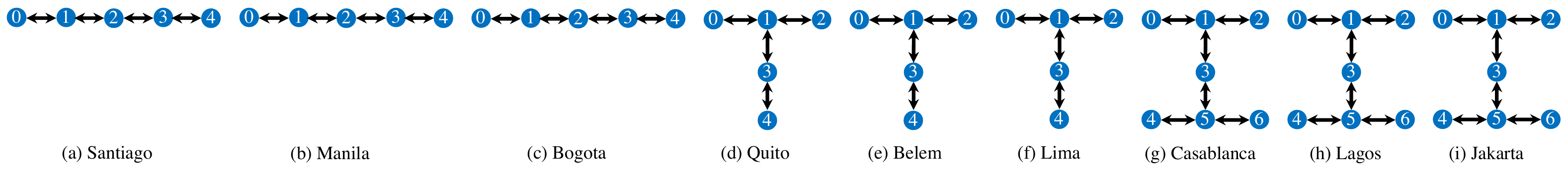}
  \caption{\small The $9$ IBM Q machines (backends) used in the evaluation. The figure shows the qubits and physical  topologies for each backend.  The backends can be grouped according to their graph topology: Line (left), T-shaped (center), and H-shaped (right). These are represented in text as $L_5$ backends, $T_5$ backends, and $H_7$ backends. The fingerprinting circuits are mapped onto these topologies, or subgraphs of the topologies if not all qubits are used.}
  \label{fig:ibmq}
\end{figure*}

\begin{figure*}[t!]
  \includegraphics[width=10cm]{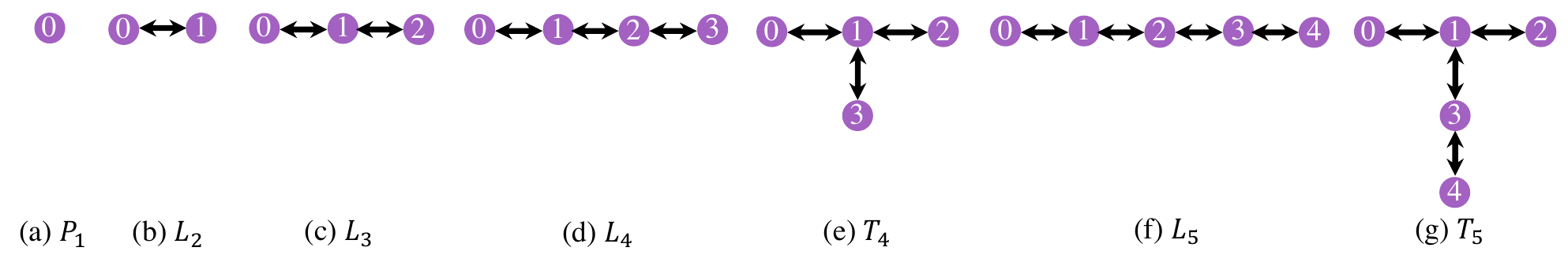}
  \caption{\small Topologies of the $7$ tomography circuits used in the evaluation. These represent attackers $\mathcal{A}$ circuits. These circuits are mapped onto the physical topologies of the backends shown in Figure~\ref{fig:ibmq}, by the provider $\mathcal{P}$.}
  \label{fig:ibmq_tomography}
\end{figure*}

\subsection{Threat Model} \label{subsec:threat_model}

The fingerprinting method proposed by this paper is based on the usual enrollment-inference paradigm. The threat model consists of an attacker $\mathcal{A}$ and a cloud provider $\mathcal{P}$. The cloud provider $\mathcal{P}$ manages $k$ devices $D_1, \ldots D_k$, where the topology of each device $D_i$, $i \in \{1, \ldots, k\}$ is described by a directed graph $G_i = (V_i, E_i)$.  As discussed later, in this work $k = 9$ due to the $9$ IBM Q backend machines used. Vertices $V_i$ represent the superconducting qubits in each physical quantum computer, and the edges $E_i$ represent couplings between the qubits. We assume that all $k$ devices support a common set of single- and two-qubit gates.

The topology of each device acts as a constraint for the placement of single- and two-qubit gates. Specifically, suppose some device $D_i$ with topology $G_i$ supports single-qubit operation $U_1(\cdot)$ and two-qubit operation $U_2(\cdot, \cdot)$. $U_1$ is supported on a qubit $q$ if and only if $q \in V_i$, and $U_2$ is supported on a pair of qubits $p, q$ if and only if $(p, q) \in E_i$. Hence for a quantum circuit $C$, we may define its \textit{topology dependency} $G_C$ as the minimal graph that supports all single- and two-qubit operations in $C$.

The task of the attacker $\mathcal{A}$ is to gather one or more full-device fingerprints for each $G_i$ during enrollment. Later in the inference phase, based on this information, $\mathcal{A}$ will collect new fingerprint data and attempt to match it to specific devices or localities on specific devices.

The fingerprinting circuits we use are implemented in unmodified Qiskit provided by IBM Q. We assume (as is done today) that the cloud provider $\mathcal{P}$ runs attacker's circuits without any modifications.  We do not require pulse-level control either, only user-level access. Obfuscation of circuits to hide the attacks or recompilation of circuits to prevent the attacks, among others, are orthogonal and future research topics.

\subsection{Fingerprint Enrollment} \label{subsec:enrollment}

During the enrollment stage, $\mathcal{A}$ gains knowledge of each $G_i$ by running a set of full-device idle tomography circuits on $D_i$ subject to single- and two-qubit drive. Given a topology $G = (V, E)$, the idle tomography experiments required for computing fingerprint $f(G)$ are categorized as follows:
\begin{itemize}
    \item For each qubit $q \in V$, use $q$ for single-qubit drive and set $V \setminus \{q\}$ as spectator qubits used for measuring crosstalk.
    \item For each coupling $(p, q) \in E$, use $p, q$ for two-qubit drive and set $V \setminus \{p, q\}$ as spectator qubits used for measuring~crosstalk.
    \item Perform further two sets of control-group experiments with all qubits as spectator qubits and idle gate delays corresponding to the single- and two-qubit drive respectively.
\end{itemize}

The circuit measurement results are analyzed with the \texttt{pyGSTi} package~\cite{nielsen2020probing}. The resultant weight-1 and weight-2 Hamiltonian, stochastic and affine errors constitute the fingerprint $f(G)$. When $G$ corresponds to a full-device topology $G_i$, these error rates constitute a full-device fingerprint $f(G_i) \in \mathbb{R}^n$ of $G_i$ for some $n \in \mathbb{Z}$. In practice, $\mathcal{A}$ may iterate through all devices multiple times. Each complete pass constitutes a \textit{batch}, and after some $l$ batches taken at distinct points in time, $\mathcal{A}$ acquires the fingerprint set $F := \{f_j(G_i) \mid j \in \{1, \ldots, l\}, i \in \{1, \ldots, k\}\}$.

\begin{figure*}[t!]
    \centering
    \includegraphics[trim={0 0.1cm 0 1.0cm},clip,width=0.9\textwidth]{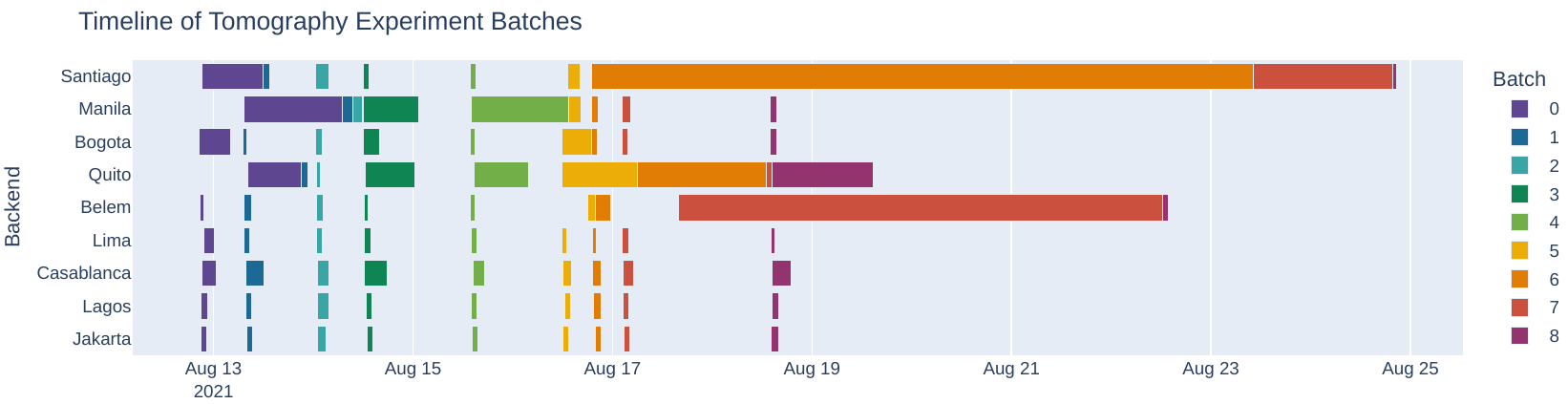}
    \caption{\small Experiment timeline showing when fingerprint data was collected from the 9 backends. The fingerprint collection was divided into 9 batches over 12 days. For each backend, each block displays the time between the invocation of the first circuit and the completion of the last circuit in a batch. Some batches took a significant amount of time to finish due to other backend users in the fair-share queue and device maintenance on IBM's end.}
    \label{fig:timeline}
\end{figure*}

\subsection{Fingerprint Matching}

During the inference stage, $\mathcal{A}$ requests $\mathcal{P}$ to run a circuit with topology dependency $G' = (V', E')$. If there exists some topology $\hat{G} \in \{G_1, \ldots, G_k\}$ such that $G'$ is isomorphic to a subgraph $G^*$ of $\hat{G}$, $G'$ is determined to be \textit{satisfiable}. In this case, $\mathcal{P}$ selects such $\hat{G} \supset G^*$, allocates the \textit{locality} $G^*$ on $\hat{G}$, and establishes a bijective \textit{embedding} $\kappa: G' \to G^*$. The objective of $\mathcal{A}$ is then to infer $G^*$ and $\phi$. Hence $\mathcal{A}$ proceeds to run idle tomography circuits on $G'$, while $\mathcal{P}$ translates the circuits on $G'$ to circuits on $G^*$ via $\phi$, without exposing $G^*$ and $\hat{G}$ to $\mathcal{A}$. After $\mathcal{P}$ returns the circuit measurement results to $\mathcal{A}$, the latter computes a fingerprint $f(G')$.

$\mathcal{A}$ then attempts to infer $G^*$ from $f(G')$. To do this, $\mathcal{A}$ iterates through $G_1, \ldots G_k$ and identifies the set $\{G'' \mid G'' \cong G'\}$ of subgraphs isomorphic to $G'$. For each $G''$ a subgraph of some $G_i$, $\mathcal{A}$ utilizes $\{f_j(G_i) \mid j \in \{1, \ldots, l\}\} \subset F$ to compute $F_{G''} := \{f_j(G'') \mid j \in \{1, \ldots, l\}\}$. The set $F_{G''}$ for each $G''$ is then used as a training set to train a classifier $\mathcal{C}_{G'}$ sensitive to fingerprints of isomorphisms of $G'$. Finally, $\mathcal{A}$ takes the prediction of $\mathcal{C}_{G'}$ on $f(G')$ as the inferred locality. The inference is correct if and only if $\mathcal{C}_{G'}(f(G')) = G^*$.

%% file: evaluation_setup.tex
\section{Evaluation Setup}

We evaluate the effectiveness of the fingerprinting scheme on $9$ IBM~Q machines (backends) shown in Figure \ref{fig:ibmq}. The machines were used to run $9$ batches of tomography experiments over 12 days. For idle tomography, we examine the idle sequence lengths 1, 2, 4 and 8. All circuits are run and measured for 2048 shots. Each batch generates one full-device fingerprint for each backend. Figure \ref{fig:timeline} shows the timeline of the tomography experiments. It is important to run multiple batches over at least few days, as there are periodic event when IBM calibrates the backends.  Our evaluation captures the calibration events and shows fingerprinting works across calibrations.

Note, the individual fingerprinting measurements can be gathered relatively quickly. Generating one full-device fingerprint takes less than an hour on 5-qubit devices, and less than two hours on 7-qubit devices. The time it takes to fingerprint a proper subgraph of each backend is less than that for the full-device fingerprint, and is dependent on the size of the specific subgraph in question.

\subsection{Choice of Subgraph Topologies}

We evaluated $7$ different subgraph topologies: $P_1$, $L_2$, $L_3$, $L_4$, $T_4$, $L_5$, $T_5$, as is shown in Figure \ref{fig:ibmq_tomography}. Note that $L_5$ and $T_5$ are themselves full-device topologies (i.e. they occupy whole backend on $L_5$ and $T_5$ devices respectively). These topologies are selected because they are the only ones that are subgraphs of more than one backend topology among the $9$~devices. For each subgraph topology, we consider all of its possible embeddings across all devices. Each subgraph topology can be embedded (i.e. mappped to the physical machines) in many ways, and it is the attacker's goal to find out where their circuit was mapped.  For example there are $84$ ways for cloud provider to map an $L_3$ circuit to one of the $9$ machines.  Our fingerprinting method is first to show ability for attacker to use the IDT fingerprints to find out which of such dozens of possible embeddings the cloud provider used.

\subsection{Data Preparation} 

To limit the experimentation time, the physical measurements were done using the $L_5$ and $T_5$  full-device topologies to obtain the full-device~fingerprint.  Fingerprint data for the other subgraphs can be comptued from $L_5$ or $T_5$ data.  Given full-device fingerprint, for each subgraph topology in $P_1$, $L_2$, $L_3$, $L_4$, $T_4$, $L_5$, $T_5$, we follow the methodology described in Subsection~\ref{subsec:enrollment} to extract the fingerprints for each of its embedded localities from the full-device~fingerprint.

\subsection{Fidelity Characterization}

We characterize the fidelity of the fingerprints with two methods:

First, we compare the inter- and intra-embedding $L^2$ separation between fingerprints for each subgraph topology and detail the results in Subsection~\ref{subsec:results:distances}.

We then train a classifier for each subgraph topology from their locality-specific fingerprints. The fingerprint data is standardized and preprocessed with principal component analysis (PCA). The main classifier is a neural network that consists of a dense layer with sigmoid activation, where the number of units matches the pre-PCA feature space dimension. This is followed by a dropout layer of drop frequency $0.2$ to prevent overfitting. Finally, a dense layer with linear activation is used, where the output dimensionality matches the number of embeddings of the subgraph topology. We use categorical cross-entropy as the loss function and use the ADAM~\cite{kingma2014adam} algorithm for stochastic gradient descent. For each model, we repeatedly train sets of $100$ epochs until the loss is less than a threshold value of $0.05$. We examine the prediction performance of the model in Subsection~\ref{subsec:results:accuracy} and discuss the degradation of prediction accuracy with time in Subsection~\ref{subsec:results:degradation}.

%% file: results.tex
\section{Results}

This section shows the fingerprinting results.

\subsection{\bf Inter- and Intra-Embedding Distances} \label{subsec:results:distances}

\begin{figure}[t]
    \centering
    \includegraphics[trim={0 0.0cm 0 1.0cm},clip,width=0.9\linewidth]{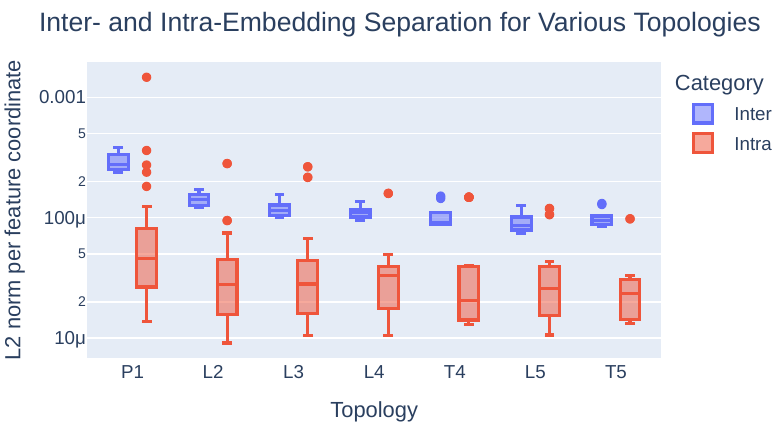}
    \caption{\small Distributions of inter- and intra-embedding separation for various topologies across all batches. The separation for two fingerprints are calculated as their $L^2$ distance divided by their feature-space dimensionality. This metric ensures comparability between different subgraph topologies. Note that the $y$ axis is log-scaled.}
    \label{fig:intra_inter}
\end{figure}

\begin{figure*}[t]
    \centering
    \includegraphics[trim={0 0.0cm 0 1.0cm},clip,width=0.76\textwidth]{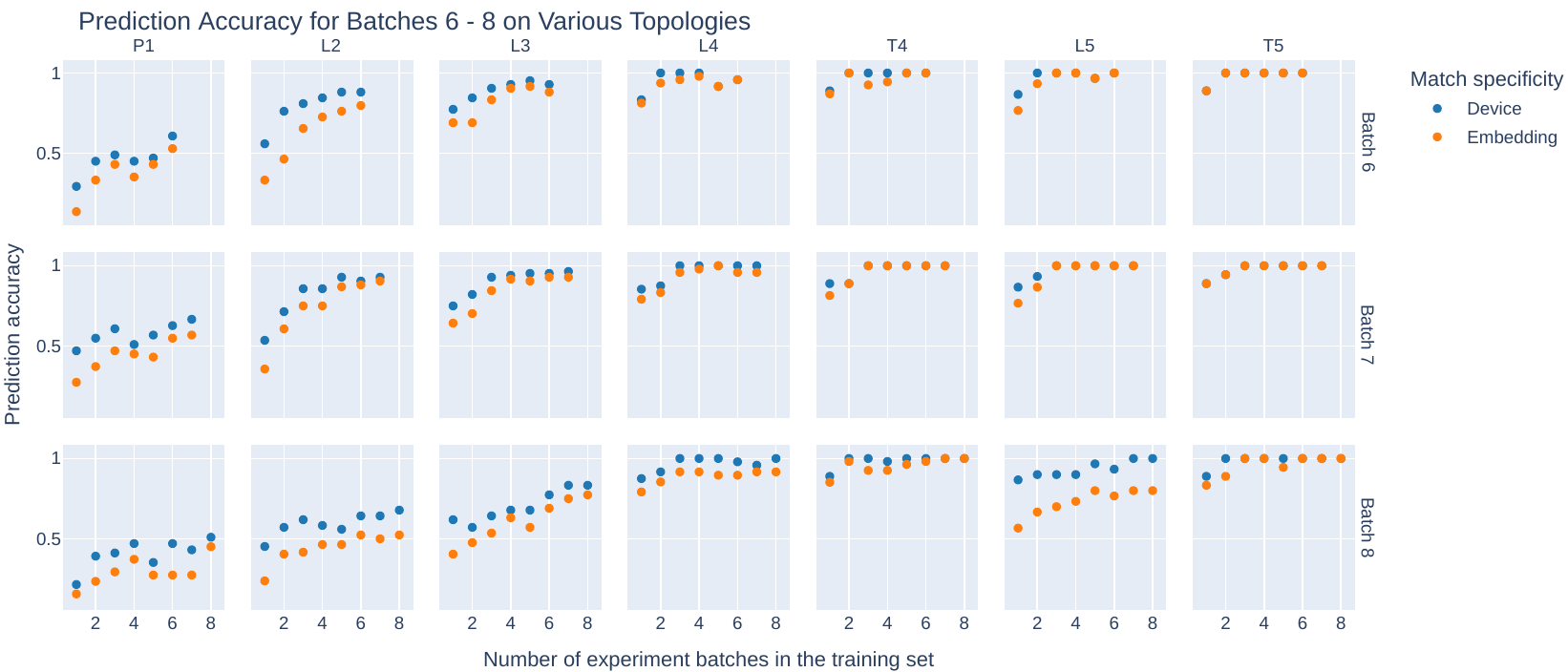}
    \caption{\small Device- and locality-specific prediction accuracy on the last 3 batches, when the training set contains the first $n$ batches for $n \in [1, 8]$.}
    \label{fig:batches_6_8_accuracy}
\end{figure*}

\begin{figure}[t]
    \centering
    \includegraphics[trim={0 0.0cm 0 1.0cm},clip,width=0.76\linewidth]{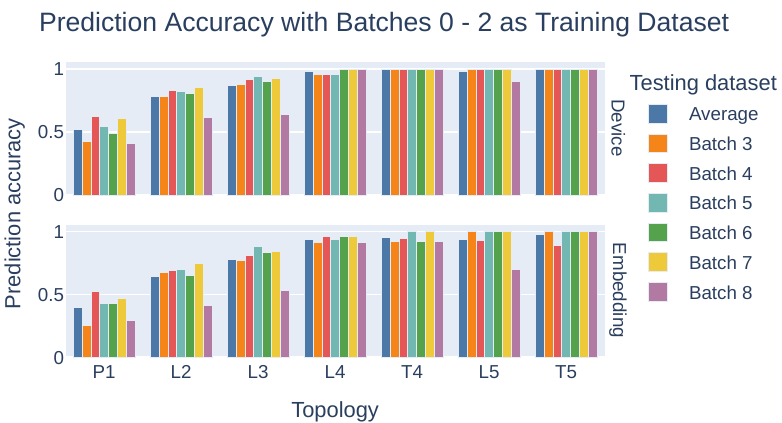}
    \caption{\small Device- and embedding-specific prediction accuracy for classifiers trained on batches 0 to 2.}
    \label{fig:3_batches_accuracy}
\end{figure}

\begin{figure}[t]
    \centering
    \includegraphics[trim={0 0.0cm 0 0.7cm},clip,width=0.76\linewidth]{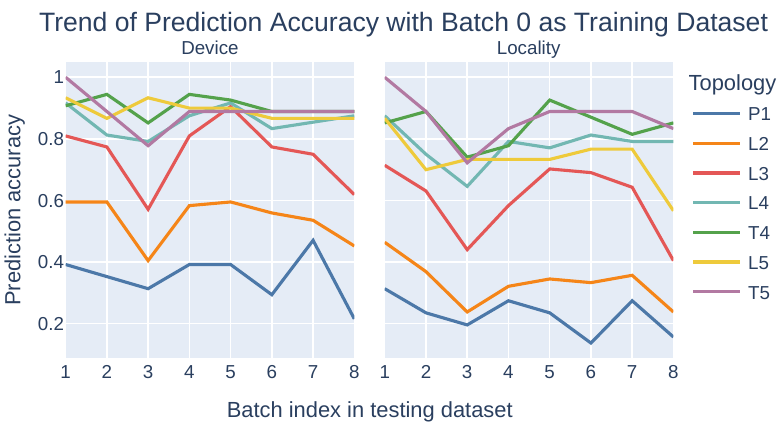}
    \caption{\small Trend of device- and locality-specific prediction accuracy for classifiers trained on batch 0.}
    \label{fig:accuracy_trend}
\end{figure}

A comparison of inter- and intra-embedding distances is shown in Figure \ref{fig:intra_inter}. Each inter-embedding distance is calculated from a pair of fingerprints on distinct embeddings in one batch, and each intra-embedding distance describes the distance between two fingerprints of the same embedding, but taken in two distinct batches. To facilitate comparison, the distance values are normalized to the feature-space dimensionality of each subgraph topology. Observe that for all topologies, there exists clear separation between the two~distributions, showing that the fingerprints clearly can distinguish between same and different embeddings.

\subsection{\bf Prediction Accuracy} \label{subsec:results:accuracy}

To test the prediction accuracy, we vary the size of the training set and examine the resultant prediction accuracy for various subgraph topologies on two levels of specificity:
\begin{itemize}
    \item Device-specific. A prediction is considered correct if the predicted locality exists on the same device as the true locality.
    \item Embedding-specific. A prediction is considered correct if and only if the predicted locality exactly matches the true locality.
\end{itemize}
Shown in Figure \ref{fig:batches_6_8_accuracy}, as the number of batches in the training set increases, the prediction accuracy values increase substantially. Apart from an outlier of $L_5$ in batch 8, complex subgraph topologies are easier to pinpoint regardless of specificity. Observe that accuracy values for most of the 4-qubit and 5-qubit topologies reach $\sim100\%$ when at least 3 batches are in the training set.

The above justifies the selection of 3 batches (0 to 2) as the training set. Under this setting, the performance on various testing sets is shown in Figure \ref{fig:3_batches_accuracy}. Among $L_4$, $T_4$, $L_5$ and $T_5$ topologies, the average device- and locality-specific prediction accuracy is $99.1 \%$ and $95.3 \%$ respectively.

\subsection{\bf Accuracy Degradation over Time} \label{subsec:results:degradation}

Figure \ref{fig:accuracy_trend} displays the trend of accuracy degradation relative to testing datasets. Note that a higher batch index in the testing dataset corresponds to a longer period in time between training and testing. Observe that in general, subgraph topologies that consist of at least $4$ qubits display no significant degradation in prediction accuracy. On the other hand, accuracy values for smaller topologies degrade moderately over time. This result demonstrates resilience and stability of the proposed fingerprint scheme, as classifiers trained on only one batch of fingerprints remain effective over a duration of at least 12 days, especially for larger subgraph topologies.

%% file: conclusion.tex
\section{Conclusion}

This work demonstrated the new threat of fingerprinting of quantum computers using crosstalk, and evaluated the approach on IBM Q cloud-based quantum computers. The device- and location-specific fingerprinting were demonstrated with accuracy to be 99.1\% and 95.3\%, respectively. We showed excellent fingerprinting abilities across many machines and across different calibration periods.

%% file: acknowledgement.tex
\section*{Acknowledgements}

The authors would like to thank IBM and Yale University for providing access to IBM's superconducting devices. This work was supported by NSF grant \nsf{1901901}. Shuwen Deng was supported through the Google PhD Fellowship.